\documentclass[11pt]{article}
\pdfoutput=1

\usepackage{bbm}

\usepackage{jheppub}
\usepackage{bigints}
\usepackage{rotating}

\usepackage{tabularx}

\usepackage{tikz}
\usepackage{fancybox}
 
\usepackage{tikz-cd}
\usepackage{ae}
\usepackage{url}

\usepackage{tabularx}

\usepackage{graphicx}
\usepackage{dsfont}
\usepackage{setspace}
\usepackage{amsfonts,amsmath,amsthm,amssymb,amsbsy}
\usepackage{longtable}
\usepackage{array}
\usepackage{color}
\usepackage{needspace}
\usepackage[numbers]{natbib}
\usepackage{colortbl}

\usepackage{enumitem}

\usepackage{fancyhdr}

\usepackage{tabularx}

\usepackage{graphicx}
\usepackage{dsfont}
\usepackage{setspace}
\usepackage{amsfonts,amsmath,amsthm,amssymb,amsbsy}
\usepackage{longtable}
\usepackage[latin1]{inputenc}
\usepackage{array}
\usepackage{color}
\usepackage{needspace}
\usepackage[numbers]{natbib}
\usepackage{colortbl}

\usepackage{xcolor}
\definecolor{colourO}{HTML}{4FEEBB}
\definecolor{colourL}{HTML}{5DB197}
\definecolor{colourG}{HTML}{3E7CEF}
\definecolor{colourK}{HTML}{3535E3}

\usepackage{enumitem}

\usepackage{fancyhdr}

\title{The Grassmannian for Celestial Superamplitudes}

\author[1,2]{Livia Ferro}\emailAdd{livia.ferro@lmu.de}
\author[2]{and Robert Moerman}\emailAdd{r.moerman@herts.ac.uk}

\affiliation[1]{Arnold--Sommerfeld--Center for Theoretical Physics,\\ Ludwig--Maximilians--Universit\"at, \\ Theresienstra\ss e 37, 80333 M\"unchen, Germany }
\affiliation[2]{Department of Physics, Astronomy and Mathematics, \\ University of Hertfordshire, \\  Hatfield, Hertfordshire, AL10 9AB, United Kingdom}

\abstract{Recently, scattering amplitudes  in four-dimensional Minkowski spacetime have been interpreted  as conformal correlation functions on the two-dimensional celestial sphere, the so-called celestial amplitudes. 
In this note we consider tree-level scattering amplitudes in $\mathcal{N}=4$ super Yang-Mills theory and present a Grassmannian formulation of their celestial counterparts. This result paves the way towards a geometric picture for celestial superamplitudes, in the spirit of positive geometries. }
\begin{document}
\begin{flushright}
{\small LMU-ASC 26/21}
\end{flushright}
\maketitle



\section{Introduction}

In the traditional momentum space, amplitudes are specified by asymptotic wave functions, which transform simply under spacetime translations, and describe transitions between momentum eigenstates. But what if four-dimensional Minkowski space is not the right space to see all their properties?
Very recently, attention has surged towards \emph{celestial amplitudes}. For the massless case, celestial amplitudes are the Mellin transform of scattering amplitudes in four-dimensional Minkowski space and are conformal correlation functions on the two-dimensional celestial sphere, i.e.\ the null infinity of four-dimensional Minkowski space. This map was proposed in \cite{Cheung:2016iub}, and explicitly applied to three and four-point tree-level gluon amplitudes in \cite{Pasterski:2017ylz} for the first time. It stems from  the observation that there is a basis which represents massless spin $s$ free fields in four-dimensional Minkowski space as spin $s$ conformal primaries on the two-sphere. Indeed, the Lorentz group  $SO(1,3)$ is mapped to the two-dimensional conformal group $SL(2,\mathbb{C})$: therefore, scattering amplitudes (\emph{i.e.}\ momentum eigenstates) in four-dimensional Minkowski space can be interpreted as two-dimensional conformal correlators (\emph{i.e.}\ boosts eigenstates) of primary fields. 
Since its inception, a lot of progress has been made: in the computation of the Mellin transform of various tree-level gluon amplitudes \cite{Schreiber:2017jsr}, of all-loop four-point amplitudes \cite{Gonzalez:2020tpi}, and of string amplitudes \cite{Stieberger:2018edy}; in the study of symmetries and  soft theorems \cite{Donnay:2018neh, Stieberger:2018onx,Himwich:2019dug, Fan:2019emx, Pate:2019mfs, Adamo:2019ipt, Nandan:2019jas, Puhm:2019zbl, Pate:2019lpp,Guevara:2019ypd, Fotopoulos:2019vac}; 
in extensions to supersymmetric theories \cite{Fotopoulos:2020bqj,Jiang:2021xzy,Brandhuber:2021nez,Hu:2021lrx}.

On one side, understanding the precise nature of the two-dimensional conformal field theory on the celestial sphere would enable a holographic description of spacetime. On the other, celestial amplitudes provide us with yet another powerful tool for exploring the mathematical properties of ordinary amplitudes by using our knowledge of conformal field theories. However,  our understanding is still preliminary and more work is necessary to advance it to the same level as, or even beyond, our understanding of ordinary amplitudes. We have only started to walk the same path which we travelled for understanding ordinary amplitudes -- analytic properties, factorization, symmetries  -- nevertheless, we do not know anything about their geometric structure. The natural question which arises is whether we can learn more about scattering amplitudes, and therefore further constrain the S-matrix,  by understanding of the conformal field theory on the celestial sphere.  Moreover, does a positive geometry \cite{Arkani-Hamed:2017tmz} for celestial amplitudes exist?

In this paper we start to fill this gap by presenting a Grassmannian formulation of celestial superamplitudes. A first look at Grassmannian formulae for amplitudes has been presented in \cite{Sharma:2021gcz}, where  tree-level gluon amplitudes have been considered. We want to go beyond this and consider the  supersymmetric version. 
Indeed, if we want to retrace the path we have covered for ordinary amplitudes in momentum space, this is the first step towards the definition of (positive) geometries for celestial amplitudes. In particular, the results of \cite{Arkani-Hamed:2009ljj,Mason:2009qx}, which presented  tree-level scattering amplitudes in $\mathcal{N}=4$ super Yang-Mills (sYM) theory in terms of integrals over a Grassmannian space, paved the way for the discovery of the amplituhedron \cite{Arkani-Hamed:2013jha} and, more generally, positive geometries underlying scattering amplitudes in various theories.
Therefore, we consider scattering amplitudes in $\mathcal{N}=4$ sYM. This theory enjoys a high amount of symmetry, the infinite-dimensional Yangian symmetry \cite{Drummond:2009fd}, which is beautifully realized in the Grassmannian formulation \cite{Drummond:2010qh,Drummond:2010uq}. This feature of the theory makes it simpler to explore and therefore a useful playground to investigate properties of amplitudes. 
Moreover, while all previous explicit results for celestial amplitudes were either worked out on a case-by-case basis or resulted in very involved expressions, a Grassmannian formula is  usually very neat and renders some properties  more manifest.
To this end, in this note 
we perform the Mellin transform of the Grassmannian integral representation of the $n$-point  helicity $k$ tree-level superamplitude and thereby find its celestial counterpart. This is the long sought-after Grassmannian formulation of celestial superamplitudes and the starting point for a plethora of investigations which are expected to bring us novel information both on the holography side and on the ordinary amplitudes side.

This note is organised as follows. In the next section we review some definitions which will be useful for presenting our main result. In particular, we will review celestial amplitudes and the  Grassmannian formulation of scattering amplitudes in $\mathcal{N}=4$ sYM. Thereafter we perform the Mellin transform of  the Grassmannian formulation and find its  celestial version. We then present some examples.


\section{Definitions}

In this section we briefly review some of the definitions that will be needed for the main result of this note. In particular, we review the celestial (super)amplitudes and the Grassmannian formulation of scattering amplitudes in $\mathcal{N}=4$ sYM.

\subsection{Celestial Amplitudes}
\label{sec:CelAmpl}

In the following we will consider tree-level scattering amplitudes of massless particles. The massless momenta can be then written in terms of the spinor-helicity variables $p^{\alpha\dot\alpha} = \lambda^\alpha \tilde\lambda^{\dot\alpha}$, $\alpha,\dot{\alpha}=1,2$. To map the amplitudes from Minkowski space to the celestial sphere, we first need to introduce celestial coordinates and parametrize the massless momenta in the following way:
\begin{equation}
p^{\alpha\dot\alpha}=\epsilon \,w\, k^{\alpha\dot\alpha} \,,
\end{equation}
where $w$ is  the angular frequency associated to the energy of the particle, conventionally called ``energy", and $k^{\alpha\dot\alpha}=\xi^\alpha \tilde\xi^{\dot\alpha}$ with
\begin{equation}
\xi^\alpha =\begin{pmatrix}
	z\\  1  
	\end{pmatrix}, \qquad 
	\tilde\xi^{\dot\alpha}=\begin{pmatrix}
	\bar{z} \\ 1
	\end{pmatrix} \,,
\end{equation}
where $z,\bar z$ are the coordinates on the celestial sphere. $\epsilon$ is a sign which depends on whether the particle is incoming or outgoing. We parametrise the spinors as
\begin{eqnarray}
\label{lambdas}
	\lambda^\alpha = \epsilon \sqrt{w } \begin{pmatrix}
	z\\  1  
	\end{pmatrix} = \epsilon \sqrt{w } \,\xi^\alpha,
	&& \tilde{\lambda}^{\dot{\alpha}}  = \sqrt{ w} \begin{pmatrix}
	\bar{z} \\ 1
	\end{pmatrix} = \sqrt{ w}\, \tilde\xi^{\dot\alpha} \,.
\end{eqnarray}
Since we are working with superamplitudes, we also have Grassmann variables $\eta^A$s. In the language of celestial amplitudes, they will be called $\tilde\tau^A$ \cite{Brandhuber:2021nez}.
After performing this change of variables, the massless scattering superamplitude is mapped on the celestial sphere via a Mellin transform \cite{Jiang:2021xzy, Brandhuber:2021nez}: 
	\begin{eqnarray}
		\label{eq1}
		\tilde{A}_{n,k}(\{\Delta_i, z_i, \bar{z}_i,\tilde\tau_i\}) =  \prod_{i=1 }^{n} \int_0^\infty d w_i w_i^{\Delta_i-1} A_{n,k}(\{w_i, z_i, \bar{z}_i,\tilde\tau_i\}) \,,
 	\end{eqnarray}
where  the celestial superamplitude transforms as a two-dimensional conformal correlator on the celestial sphere with weights $\Delta_i$ . 
We end this section by remarking  that, under conformal transformations, we have the following behaviour
\begin{equation}
\label{conftrans}
w \to w' =|cz+d|^2 w \,, \quad z\to z'=\frac{a z+b}{c z+d}\,,\quad\bar z\to \bar z'=\frac{\bar a \bar z+\bar b}{\bar c \bar z+\bar d}\,,\quad \tilde\tau^A \to \tilde\tau'^A =\frac{(cz+d)^{1/2}}{(\bar{c}\bar{z}+\bar{d})^{1/2}}\tilde\tau^A\,,
\end{equation}
with $a,b,c,d\in\mathbb{C}$ and $a d-b c=1$.

\subsection{Grassmannian Integrals}
\label{sec:GrAmpl}

Let us  now review the formulation of scattering amplitudes in $\mathcal{N}=4$ sYM in terms of Grassmannian integrals.
In \cite{Arkani-Hamed:2009ljj,Mason:2009qx} the leading singularities of the $\mathcal{N}=4$ sYM 
N${^{k-2}}$MHV $n$-point amplitudes  were described by an integral over the space of $k$-planes in $n$ dimensions, the Grassmannian $G(k,n)$, along suitable closed
contours. 
This arose from the observation that momentum conservation, which is a quadratic constraint in the spinor-helicity space, can be linearised by using auxiliary spaces. 
Indeed, we can introduce an auxiliary $k$-plane in $n$-dimensions, $C=(c_{ai})$, such that
\begin{equation}
C \cdot \tilde\lambda=0\,, \qquad \lambda\cdot C^\perp=0 \,,
\end{equation}
where $C^\perp$ is the orthogonal complement of $C$. In this way, the condition  $\sum_{i=1}^n \lambda_{i}^{\alpha} \widetilde\lambda_{i}^{\dot \alpha}=0$ is linearised.
The tree-level amplitudes in spinor-helicity space can therefore be written as
\begin{equation}
\label{Grampl}
A_{n,k} = \int_\gamma d\Omega_{n,k}\prod_{\dot{a}=1}^k 
\delta^{2}\left(\sum_{i=1}^n c_{\dot{a}i} \tilde\lambda_i \right) \prod_{a=k+1}^n 
\delta^{2}\left(\sum_{i=1}^n {c_{ai}^\perp} \lambda_i \right)  \prod_{\dot{a}=1}^k  \delta^{4}\left(\sum_{i=1}^n c_{\dot{a}i} \eta_i \right) \,,
\end{equation}
where $\gamma$ is a closed contour.
The measure is 
\begin{equation}
 d\Omega_{n,k} = \frac{\prod_{\dot{a},i} dc_{\dot{a}i}}{GL(k)(1\ldots k)(2\ldots k+1)\ldots(n \ldots n+k-1)} 
\end{equation}
where the denominator consists of the cyclic product of the minors $\mathcal{M}_i=(i\, i+1...i+k-1)$, \emph{i.e.} the determinants of $(k \times k)$ submatrices of the  matrix $C$. 
The contour $\gamma$ can be determined by using \emph{e.g.}~the Britto-Cachazo-Feng-Witten (BCFW) recursion relations  \cite{Britto:2004ap,Britto:2005fq}.


\section{Grassmannian on the Celestial Sphere}
\label{sec:GrCelestial}

We are now ready to consider our  scattering superamplitude in the Grassmannian formulation \eqref{Grampl} and perform the Mellin transform. Despite the simplicity of this idea, we will soon discover that  in the process a few tricks need to be used in order to find a neat result.
We start by using the $GL(k)$ redundancy to write $C\in G(k,n)$ as
\begin{align}
C_{\dot{a}\dot{b}}=\delta_{\dot{a}\dot{b}}\,,\qquad C_{\dot{a}b}=c_{\dot{a}b}\,.
\end{align} 
With this particular choice, the bosonic delta-functions in \eqref{Grampl} read
\begin{align}
\delta^{k\times 2}(C \cdot\tilde{\lambda})=\prod_{\dot{a}}\delta^2(\tilde{\lambda}_{\dot{a}}+\sum_{b}c_{\dot{a}b}\tilde{\lambda}_b)\,,\qquad
\delta^{2\times(n-k)}(\lambda \cdot C^\perp)=\prod_{a}\delta^2(\lambda_{a}-\sum_{\dot{b}}c_{\dot{b}a}\lambda_{\dot{b}})\,.
\end{align}
If we now move to celestial coordinates, these delta-functions will contain square roots of the form $\sqrt{\omega_i}$ introduced via \eqref{lambdas}. To remove these and render the computation easier, we utilize the trick presented in \cite{Sharma:2021gcz} and use the little group scaling transformations
\begin{align}
\lambda_i\to t_i\lambda_i\,,\qquad\tilde{\lambda}_i\to t_i^{-1}\tilde{\lambda}_i\,,\qquad \tilde{\eta}_i\to t_i^{-1}\tilde{\eta}_i\,,
\end{align}
which are reflected on the superamplitude as $A_n\to t_i^{2}\, A_n$. In particular, we choose different forms for the scalings of different particles:
\begin{align}
t_{\dot{a}}= w_{\dot{a}}^{-1/2}\,, \,\dot{a}=1,\ldots,k\,, \qquad t_{a}= w_{a}^{1/2} \,, \, a=k+1,\ldots,n\,.
\end{align}
After performing this little group transformation, we find that
\begin{align}
\nonumber &\tilde{A}_{n,k}=\int_{\gamma} d\Omega_{n,k} \left(\prod_{i=1}^{n}\int_{0}^{\infty}\frac{dw_i}{w_i}w_i^{\Delta_i}\right)\prod_{\dot{a}}w_{\dot{a}}^{-1}\prod_{a}w_{a}\\
&\cdot \prod_{\dot{a}}\delta^2\Big(w_{\dot{a}}\tilde{\xi}_{\dot{a}}+\sum_b c_{\dot{a}b}\tilde{\xi}_{b}\Big)
\prod_{a}\delta^2\Big(\epsilon_{a}w_{a}\xi_{a}-\sum_{\dot{b}} c_{\dot{b}a}\epsilon_{\dot{b}}\xi_{\dot{b}}\Big)
\prod_{\dot{a}}\delta^4\Big(\sqrt{w_{\dot{a}}}\tilde{\tau}_{\dot{a}}+\sum_b c_{\dot{a}b}\frac{\tilde{\tau}_{b}}{\sqrt{w_b}}\Big)\,,
\end{align}
where now no square root is present in the bosonic delta-functions.
To illuminate the transformational properties of $\tilde{A}_{n,k}$, it will prove useful to introduce auxiliary parameters $p_i$ and $\bar{p}_i$ for each particle $i$, where $\bar{p}_i=(p_i)^\ast$ or both parameters are real and independent. Rewriting $w_i$ as 
\begin{align}
w_i\equiv (p_i\bar{p}_i)^2y_i\,,
\end{align}
we find
\begin{align}
\nonumber \tilde{A}_{n,k}&=\prod_{i=1}^n\left[(p_i\bar{p}_i)^{2}\right]^{\Delta_i}
\prod_{\dot{a}}{\frac{p_{\dot{a}}^2}{\bar{p}_{\dot{a}}^6}}
\prod_{a}\frac{1}{(p_a\bar{p}_{a})^2}\int_{\gamma} d\Omega_{n,k}\left(\prod_{i=1}^{n}\int_{0}^{\infty}{dy_i}y_i^{\Delta_i}\right)\\
\nonumber &\cdot \prod_{\dot{a}}\delta^2\Big(y_{\dot{a}}\tilde{\xi}_{\dot{a}}+\sum_b c_{\dot{a}b}\frac{\tilde{\xi}_{b}}{(p_{\dot{a}}\bar{p}_{\dot{a}})^2}\Big)
\prod_{a}\delta^2\Big(\epsilon_{a}y_{a}\xi_{a}-\sum_{\dot{b}} c_{\dot{b}a}\frac{\epsilon_{\dot{b}}\xi_{\dot{b}}}{(p_{a}\bar{p}_{a})^2}\Big)
\\
&\cdot \prod_{\dot{a}}\delta^4\Big(\tau_{\dot{a}}+\sum_b \frac{c_{\dot{a}b}}{(p_{\dot{a}}\bar{p}_{b})^2}\frac{\tau_{b}}{\sqrt{y_{\dot{a}}}\sqrt{y_b}}\Big)\,,
\end{align}
where we have defined new fermionic variables
\begin{align}
\tau_i\equiv\frac{\bar{p}_i}{p_i}\tilde{\tau}_i\,. \label{eq:mellin-tau}
\end{align}
By performing the change of variables\footnote{This change of variables leaves the Grassmannian measure $d\Omega_{n,k}$ invariant.} $c_{\dot{a}b}\to c_{\dot{a}b}(p_{\dot{a}}\bar{p}_b)^2$, the bosonic delta-functions can then be written as
\begin{alignat}{2}
&\delta^2\Big(y_{\dot{a}}\tilde{\xi}_{\dot{a}}+\sum_b c_{\dot{a}b}\frac{\tilde{\xi}_{b}}{(p_{\dot{a}}\bar{p}_{\dot{a}})^2}\Big)&&\to \bar{p}_{\dot{a}}^4\delta(y_{\dot{a}}-E_{\dot{a}})\delta\Big(\sum_bc_{\dot{a}b}(\bar{p}_{\dot{a}}\bar{p}_b)^2\bar{z}_{\dot{a}b}\Big)\,,\\
&\delta^2\Big(\epsilon_{a}y_{a}\xi_{a}-\sum_{\dot{b}} c_{\dot{b}a}\frac{\epsilon_{\dot{b}}\xi_{\dot{b}}}{(p_{a}\bar{p}_{a})^2}\Big)&&\to
p_{a}^4\delta(y_{a}-E_{a})\delta\Big(\sum_{\dot{b}}c_{\dot{b}a}(p_{\dot{b}}p_{a})^2\epsilon_{\dot{b}}\epsilon_{a}z_{\dot{b}a}\Big)\,,
\end{alignat}
where
\begin{align}
E_{\dot{a}}=-\sum_bc_{\dot{a}b}\Big(\frac{\bar{p}_b}{\bar{p}_{\dot{a}}}\Big)^2\frac{[b\tilde{\chi}]}{[\dot{a}\tilde{\chi}]}\,,\quad E_{a}=\sum_{\dot{b}}c_{\dot{b}a}\Big(\frac{p_{\dot{b}}}{p_{a}}\Big)^2\frac{\epsilon_{\dot{b}}\langle \dot{b}\chi\rangle}{\epsilon_{a}\langle a \chi\rangle}\,,
\label{eq:mellin-energies}
\end{align}
are ``energies'' as in \cite{Sharma:2021gcz} and $\chi^\alpha,\tilde{\chi}^{\dot{\alpha}}$ are auxiliary reference spinors. 
Finally, we obtain the main result of our paper: the Mellin transform of the $\mathcal{N}=4$ sYM amplitudes in the Grassmannian formulation is
\begin{align}
\label{amplcel}
\tilde{A}_{n,k}=P_{n} \cdot I_{n,k}\,,
\end{align}
where $I_{n,k}$ is an integral over the Grassmannian $G(k,n)$
\begin{align}
&\nonumber I_{n,k} =
\int_{\gamma} d\Omega_{n,k}\prod_{i=1}^{n}E_i^{\Delta_i}\mathbbm{1}_{> 0}(E_i)\\
& \cdot \prod_{\dot{a}}\delta\Big(\sum_bc_{\dot{a}b}(\bar{p}_{\dot{a}}\bar{p}_b)^2\bar{z}_{\dot{a}b}\Big)
\prod_{a}\delta\Big(\sum_{\dot{b}}c_{\dot{b}a}(p_{\dot{b}}p_{a})^2\epsilon_{\dot{b}}\epsilon_{a}z_{\dot{b}a}\Big)\prod_{\dot{a}}\delta^4\Big(\tau_{\dot{a}}+\sum_b c_{\dot{a}b}\frac{\tau_{b}}{\sqrt{E_{\dot{a}}E_b}}\Big)\,,
\label{eq:mellin-integral}
\end{align} 
and  $P_n\equiv\left(\prod_{i=1}^np_i^{2(\Delta_i+1)}\bar{p}_{i}^{2(\Delta_i-1)}\right)$. The indicator function $\mathbbm{1}_{> 0}(E_i)$ is defined to be one if $E_i$ is real and non-negative, and zero otherwise. It appears because of the integration ranges for $w$'s are the positive real line. 

If we now specify that, under a conformal transformation, $p_i$ and $\bar{p}_i$ transform according to 
\begin{align}
	p_i\to(cz_i+d)^{1/2}p_i\,,\qquad \bar{p}_i\to(\bar{c}\bar{z}_i+\bar{d})^{1/2}\bar{p}_i\,,
\end{align}
then the Grassmannian integral $I_{n,k}$ defined above is conformally invariant, while the prefactor $P_{n,k}$ transforms covariantly
\begin{align}
P_n\to \left(\prod_{i=1}^n(cz_i+d)^{\Delta_i+1}(\bar{c}\bar{z}_i+\bar{d})^{\Delta_i-1}\right)P_n\,.
\end{align}
Altogether, this ensures that $\tilde{A}_{n,k}$ transforms as expected under conformal transformations. Thus, the auxiliary parameters $p_i$ and $\bar{p}_i$ make the transformation properties of $\tilde{A}_{n,k}$ manifest. 

One could, however, make any choice for $p_i$ and $\bar{p}_i$ in order to evaluate $\tilde{A}_{n,k}$, and the result would be the same, but the transformation properties of $\tilde{A}_{n,k}$ might not be manifest. For example, choosing $p_i=\bar{p}_i=1$, we obtain the Grassmannian integral
\begin{align}
	&\nonumber A_{n,k} =
	\int_{\gamma} d\Omega_{n,k}\prod_{i=1}^{n}E_i^{\Delta_i}\mathbbm{1}_{> 0}(E_i)\\
	& \cdot \prod_{\dot{a}}\delta\Big(\sum_bc_{\dot{a}b}\bar{z}_{\dot{a}b}\Big)
	\prod_{a}\delta\Big(\sum_{\dot{b}}c_{\dot{b}a}\epsilon_{\dot{b}}\epsilon_{a}z_{\dot{b}a}\Big)\prod_{\dot{a}}\delta^4\Big(\tilde{\tau}_{\dot{a}}+\sum_b c_{\dot{a}b}\frac{\tilde{\tau}_{b}}{\sqrt{E_{\dot{a}}E_b}}\Big)\,,
	\label{eq:mellin-integral-simplified}
\end{align}
where $\tilde{\tau}_i$ are the original fermionic parameters and the energies read
\begin{align}
	E_{\dot{a}}=-\sum_bc_{\dot{a}b}\frac{[b\tilde{\chi}]}{[\dot{a}\tilde{\chi}]}\,,\quad E_{a}=\sum_{\dot{b}}c_{\dot{b}a}\frac{\epsilon_{\dot{b}}\langle \dot{b}\chi\rangle}{\epsilon_{a}\langle a \chi\rangle}\,.
	\label{eq:mellin-energies-simplified}
\end{align}
We have checked that formula \eqref{eq:mellin-energies-simplified} correctly reproduces the formula (6.13) in  \cite{Sharma:2021gcz} when projected onto pure gluon amplitudes.
	
\subsection{Examples}

\paragraph{Three-point MHV.} Let us consider the case $(n,k)=(3,2)$ with $\epsilon_1=\epsilon_2=-\epsilon_3=1$. Since this is the simplest case, we present it in full detail, with most calculations explicit.  
In Minkowski signature the on-shell three-point amplitude vanishes. Thus, we assume to be working in the $(2,2)$ split signature, where $z_i$ and $\bar{z}_i$ are real, independent variables. 
Consider the following gauge fixing
\begin{align}
C=\begin{pmatrix}
1&0&c_{13}\\
0&1&-c_{23}
\end{pmatrix},
\end{align}
where the minus sign is chosen for convenience. Consider, first, the bosonic delta-functions in \eqref{eq:mellin-integral-simplified}. They are explicitly
\begin{align}
\prod_{\dot{a}}\delta\Big(\sum_bc_{\dot{a}b}\bar{z}_{\dot{a}b}\Big)
\prod_{a}\delta\Big(\sum_{\dot{b}}c_{\dot{b}a}\epsilon_{\dot{b}}\epsilon_{a}z_{\dot{b}a}\Big)
=c_{13}^{-1}c_{23}^{-1}z_{31}^{-1}\delta(\bar{z}_{23})\delta(\bar{z}_{31})\delta\Big(c_{13}+c_{23}\frac{z_{23}}{z_{31}}\Big)\,.
\end{align} 
On the support of the last delta-function, the energies become
\begin{align}
E_1=c_{23}\frac{z_{23}}{z_{31}}\,,\qquad E_2=c_{23}\,, \qquad E_3 =c_{23}\frac{z_{21}}{z_{31}}\,,
\end{align}
and consequently, the indicator functions simplify to 
\begin{align}
\prod_{i=1}^3\mathbbm{1}_{> 0}(E_i)=\mathbbm{1}_{> 0}(c_{23})\Theta\Big(\frac{z_{21}}{z_{23}}\Big)\Theta\Big(\frac{z_{21}}{z_{31}}\Big)\,,
\end{align}
where $\Theta$ is the Heaviside step function.  Notice that $\mathbbm{1}_{> 0}(c_{23})$ restricts the integral for $c_{23}$, which was originally a complex integral, to the positive reals. The fermionic delta-functions in  \eqref{eq:mellin-integral-simplified} simplify to 
\begin{align}
\nonumber \prod_{\dot{a}}\delta^4\Big(\tau_{\dot{a}}+\sum_b c_{\dot{a}b}\frac{\tau_{b}}{\sqrt{E_{\dot{a}}E_b}}\Big)
&=\delta^4\Big(\tilde\tau_1-\sqrt{\frac{z_{23}}{z_{21}}}\tilde{\tau}_3\Big)\delta^4\Big(\tilde{\tau}_2-\sqrt{\frac{z_{31}}{z_{21}}}\tilde{\tau}_3\Big)\\
&=z_{23}^{-2}z_{31}^{-2}z_{12}^{-4}\delta^{8}(\sqrt{z_{23}}\xi_1\tilde{\tau}_1+\sqrt{z_{31}}\xi_2\tilde{\tau}_2-\sqrt{z_{21}}\xi_3\tilde{\tau}_3)\,.
\end{align}
Combining all of these simplifications, one obtains
\begin{align}
\nonumber \tilde{A}_{3,2}&=z_{12}^{\Delta_3-4}z_{23}^{\Delta_1-4}z_{31}^{-1-\Delta_{1}-\Delta_3}\delta(\bar{z}_{23})\delta(\bar{z}_{31})\Theta\Big(\frac{z_{21}}{z_{23}}\Big)\Theta\Big(\frac{z_{21}}{z_{31}}\Big)\\
&\cdot\delta^{8}(\sqrt{z_{23}}\xi_1\tilde{\tau}_1+\sqrt{z_{31}}\xi_2\tilde{\tau}_2-\sqrt{z_{21}}\xi_3\tilde{\tau}_3)\int_{0}^{\infty}\frac{dc_{23}}{c_{23}}c_{23}^{(\Delta_1+\Delta_2+\Delta_3)-3}\,.
\end{align}
Writing $\Delta_i=i\beta_i+1$ for $\beta_i\in\mathbb{R}$ and defining $\beta\equiv\beta_1+\beta_2+\beta_3$, the remaining integral simplifies to
\begin{align}
\int_{0}^{\infty}\frac{dc_{23}}{c_{23}}c_{23}^{(\Delta_1+\Delta_2+\Delta_3)-3} = 2\pi\delta(\beta)\,,
\end{align} 
and on the support of this delta-function, the three-point MHV celestial superamplitude is given by 
\begin{align}
\nonumber \tilde{A}_{3,2}&=2\pi\delta(\beta)z_{12}^{\Delta_3-4}z_{23}^{\Delta_1-4}z_{31}^{\Delta_2-4}\delta(\bar{z}_{23})\delta(\bar{z}_{31})\Theta\Big(\frac{z_{21}}{z_{23}}\Big)\Theta\Big(\frac{z_{21}}{z_{31}}\Big)\\
&\cdot\delta^{8}(\sqrt{z_{23}}\xi_1\tilde{\tau}_1+\sqrt{z_{31}}\xi_2\tilde{\tau}_2-\sqrt{z_{21}}\xi_3\tilde{\tau}_3)\,,
\end{align}
in agreement with  \cite{Jiang:2021xzy,Brandhuber:2021nez}.

The three-point $\mathrm{\overline{MHV}}$ celestial superamplitude can be calculated in a similar way, where this time the gauge-fixed $C$-matrix reads $C=\begin{pmatrix}
1&c_{13}&c_{23}
\end{pmatrix}$.
	
\paragraph{Four-point MHV.} For the case $(n,k)=(4,2)$, and $\epsilon_1=\epsilon_2=-\epsilon_3=-\epsilon_4=1$, we only present a few features as the computation is involved and tedious, nevertheless it can be performed straightforwardly. In this case the matrix $C$ reads
\begin{align}
C=\begin{pmatrix}
1&0&c_{13}&c_{14}\\
0&1&c_{23}&c_{24}
\end{pmatrix} \,.
\end{align}
The bosonic delta-functions give the following expression
\begin{align}
\nonumber &\prod_{\dot{a}}\delta\Big(\sum_bc_{\dot{a}b}\bar{z}_{\dot{a}b}\Big)
\prod_{a}\delta\Big(\sum_{\dot{b}}c_{\dot{b}a}\epsilon_{\dot{b}}\epsilon_{a}z_{\dot{b}a}\Big)\\
& =\frac{1}{c_{24}|z_{13}z_{24}|^2}\delta(r-\bar{r})
\delta\Big(c_{13}+c_{24}(r-1)\frac{|z_{24}|^2}{z_{14}\bar{z}_{23}}\Big)  \delta\Big(c_{14}+c_{24}\frac{z_{24}}{z_{14}}\Big) 
\delta\Big(c_{23}+c_{24}\frac{\bar{z}_{24}}{\bar{z}_{23}}\Big)\,,
\end{align} 
where the cross-ratios $r$ and $\bar{r}$ are, respectively, $r=\frac{z_{12}z_{34}}{z_{13}z_{24}}$ and $\bar{r}=\frac{\bar{z}_{12}\bar{z}_{34}}{\bar{z}_{13}\bar{z}_{24}}$. As already noticed in, \emph{e.g.}\ \cite{Pasterski:2017ylz},  the factor $\delta(r-\bar{r})$ ensures that in the four-particle scattering the momenta of all particles lie on the same plane. The intersection of this plane with the celestial sphere forces the cross-ratio $r$ to be real. The energies on the support of the above delta-functions become
\begin{align}
E_1=(r-1)\frac{|z_{24}|^2}{|z_{14}|^2}E_2=\tilde{c}_{24}(r-1)\frac{|z_{24}|^2}{|z_{14}|^2}\,,\quad E_3=(r-1)\frac{|z_{24}|^2}{|z_{23}|^2}E_4=\tilde{c}_{24}r\frac{|z_{24}|^2}{|z_{34}|^2}\,,
\end{align}
where $\tilde{c}_{24}=c_{24}\frac{\bar{z}_{34}}{\bar{z}_{23}}$. In particular, $E_2=\tilde{c}_{24}$. The indicator functions require these energies to be real and positive which forces $r>1$, producing $\Theta(r-1)$, and $\tilde{c}_{24}$ to be real and positive, producing $\mathbbm{1}_{> 0}(\tilde{c}_{24})$. With a little effort, one can show that the fermionic delta-functions can be written as 
\begin{align}
&\nonumber \prod_{\dot{a}}\delta^4\Big(\tilde{\tau}_{\dot{a}}+\sum_b c_{\dot{a}b}\frac{\tilde{\tau}_{b}}{\sqrt{E_{\dot{a}}E_b}}\Big)\\
&= z_{12}^{-4}\left|\frac{z_{23}z_{24}}{z_{13}z_{14}}\right|^2\delta^{8}\Big(
\xi_1\tilde{\tau}_1
+\left|\frac{z_{13}z_{14}}{z_{23}z_{24}}\right|^{1/2}\xi_2\tilde{\tau}_2
-\left|\frac{z_{12}z_{14}}{z_{23}z_{34}}\right|^{1/2}\xi_3\tilde{\tau}_3
-\left|\frac{z_{12}z_{13}}{z_{24}z_{34}}\right|^{1/2}\xi_4\tilde{\tau}_4
\Big)\,,
\end{align}
and it is clearly independent of $\tilde{c}_{24}$. The product of all energies, each raised to the power of their respective weight, evaluates to
\begin{align}
\nonumber &\prod_{i=1}^4E_i^{\Delta_i}=(-1)^{\Delta_1+\Delta_2}\tilde{c}_{24}^{\Delta_1+\Delta_2+\Delta_3+\Delta_4}\Big(\frac{z_{14}z_{23}}{z_{34}^2}\Big)^2\Big(\frac{\bar{z}_{12}^2}{\bar{z}_{14}\bar{z}_{23}}\Big)^2\\
&\cdot 
\Big(\frac{z_{13}}{\bar{z}_{42}}\Big)^{\Delta_2+\Delta_4-4}
\Big(\frac{\bar{z}_{12}}{z_{34}}\Big)^{\Delta_3+\Delta_4-4}
\Big(\frac{\bar{z}_{14}}{z_{23}}\Big)^{\Delta_2-2}
\Big(\frac{z_{14}}{\bar{z}_{23}}\Big)^{\Delta_3-2}\,.
\end{align}
Collecting all factors of $\tilde{c}_{24}$ and integrating over all $\tilde{c}_{2}$ (on the support of $\mathbbm{1}_{> 0}(\tilde{c}_{24})$) produces the delta-function
\begin{align}
\int_{0}^{\infty}\frac{d\tilde{c}_{24}}{\tilde{c}_{24}}\tilde{c}_{24}^{(\Delta_1+\Delta_2+\Delta_3+\Delta_4)-4}=2\pi\delta(\beta)\,,
\end{align}  
where, again, $\Delta_i=i\beta_i+1$ for $\beta_i\in\mathbb{R}$ and $\beta\equiv\beta_1+\beta_2+\beta_3+\beta_4$. Finally, the four-point MHV celestial superamplitude is given by 
\begin{align}
\nonumber &\tilde{A}_{4,2}=2\pi\delta(\beta)(-1)^{\Delta_1+\Delta_2}\Big(\frac{z_{13}}{\bar{z}_{42}}\Big)^{\Delta_2+\Delta_4-4}
\Big(\frac{\bar{z}_{12}}{z_{34}}\Big)^{\Delta_3+\Delta_4-4}
\Big(\frac{\bar{z}_{14}}{z_{23}}\Big)^{\Delta_2-2}
\Big(\frac{z_{14}}{\bar{z}_{23}}\Big)^{\Delta_3-2}\\
& \cdot
\frac{\delta(r-\bar{r})\Theta(r-1)}{z_{12}z_{23}z_{34}z_{41}|z_{13}|^2|z_{24}|^2}\delta^{8}\Big(
\xi_1\tilde{\tau}_1
+\left|\frac{z_{13}z_{14}}{z_{23}z_{24}}\right|^{1/2}\xi_2\tilde{\tau}_2
-\left|\frac{z_{12}z_{14}}{z_{23}z_{34}}\right|^{1/2}\xi_3\tilde{\tau}_3
-\left|\frac{z_{12}z_{13}}{z_{24}z_{34}}\right|^{1/2}\xi_4\tilde{\tau}_4
\Big)\,,
\end{align}
in agreement with \cite{Brandhuber:2021nez}.

\section{Acknowledgements}

We would like to thank Tomek \L ukowski for useful comments on the draft. L.F. thanks David Damgaard for collaboration at early stages of this project.
This work was partially funded by the Deutsche Forschungsgemeinschaft (DFG, German Research Foundation) -- Projektnummern 404358295 and 404362017.

\bibliographystyle{nb}

\bibliography{celestial_grass}

\end{document}